\pdfoutput=1
\documentclass[%
reprint,prl,
amsmath,amssymb,
aps,
]{revtex4-2}
\usepackage{xparse}

\usepackage{appendix}
\usepackage{graphicx}
				\graphicspath{ {./images/} } 
\usepackage{dcolumn}
\usepackage{bm}
\usepackage{hyperref}

\newcommand{\dd}[2]{\ensuremath{\frac{\text{d}^2 #1}{\text{d} #2^2}}}

\usepackage{slashed}
\renewcommand{\vec}[1]{\mathbf{#1}}

\usepackage{bbold}
\usepackage{comment}
\usepackage{xcolor} 

\begin{document}
\title{Moir\'e  Gravity and Cosmology} 

	\author{Alireza Parhizkar}
\affiliation{Joint Quantum Institute, University of Maryland, College Park, MD 20742, USA}
\author{Victor Galitski}
\affiliation{Joint Quantum Institute, University of Maryland, College Park, MD 20742, USA}

\date{
    \today
}

\begin{abstract}
The  vacuum catastrophe is a fundamental puzzle, where the observed scales of the cosmological constant are many orders of magnitude smaller than the natural scales expected in the theory. This work proposes a new ``bi-world'' construction that may offer an insight into the cosmological constant problem. The model generally includes a $(3+1)$-dimensional manifold with two different geometries and matter fields residing on them.
In contrast to bimetric or massive theories of gravity, classically, and in the vacuum limit, the model reduces to two massless gravity theories, thereby living ghost-free.
The diffeomorphism invariance and causality highly constrain the two metrics to be conformally related, $\eta_{\mu \nu} = \phi^2 g_{\mu \nu}$.  This reduces the theory to a standard single-world description, but introduces a new inherently geometrical ``moir{\'e} field,'' $\phi$. Interestingly, the moir{\'e} field has the character of both a dilaton  and Higgs field familiar in the conventional theory.   Integrating out the moir{\'e} field naturally gives rise to the Starobinsky action and inflationary dynamics. In the framework of the Friedmann-Lemaitre–Robertson–Walker solution, we reduce an effective action for the  moir{\'e} field to that of a particle moving in a Mexican hat potential. The equations of motion are then solved numerically and the  moir{\'e} field is shown to approach a Mexican-hat minimum in an oscillatory fashion, which is accompanied by the decay of the Hubble parameter. Under additional reasonable assumptions, the vacuum energy asymptotically approaches zero in the end of inflationary evolution. The physics presented here shares similarities with the moir{\'e} phenomena in condensed matter and elsewhere, where two similar structures superimposed upon give rise to a superstructure with low emergent energy scales compared to the native theories. 
\end{abstract}
\date{\today}

\maketitle


\textit{Introduction}---As ubiquitous phenomena,
moir\'e patterns are not hard to come by. Generally when two layers of grids combine, like overlapping fabrics or when a digital photo of a pixel screen is watched through another such screen, an additional larger pattern emerges. When the original layers are close enough, the moir\'e pattern becomes more than just an optical illusion. For example, in bilayer graphene the moir\'e pattern dictates the interlayer hopping process~\cite{EmergentScale,Balents,MoireBands,UnconventionalSC}. This concept is generalizable to continuous models as in \cite{EmergentScale} where we can define moir\'e physics to be: The emergence of large length scales, or synonymously small energy scales, due to coalition of two underlying theories with small length or large energy characteristics.

Moir\'e physics as a conceptual tool is potentially utilizable in many different contexts. Here we are going to investigate its possible presence in gravitational systems and its implication in cosmology.


\textit{Bi-world}---By definition, for a moir\'e pattern to appear, we first need two more-or-less similar systems as the underlying structures. In case of gravity those are metric systems and therefore we are combining two curved space-times, and potentially the matter fields residing on them, with each other. Such a universe that embraces two worlds is described most generally by an action $S= S_g + S_\eta + S_c + S_a$ where $S_g$ and $S_\eta$ are built only out of fields that live solely on either $g$-world or $\eta$-world, while $S_c$ contains the cross terms consisted of these fields. But there is also the possibility of \textit{amphibian} fields that live on both worlds, the physical description of which is encapsulated in $S_a$. If not through $S_c$ and $S_a$ the g-world and $\eta$-world could not talk and would have been completely independent of each other.

Naturally, we now have to integrate over all fields contributing to $S$. We usually take $S_\eta$ to be a copy of $S_g$ and then $S_c$ will contain various coupling terms between fields in g-world and their counterparts in $\eta$-world. In this fashion the path-integral looks like the following,
\begin{align}
    I = \int \mathcal{D}g_{\mu\nu} \mathcal{D}\eta_{\mu\nu} \mathcal{D}\left[ \bar{\Psi}^g,\Psi^g \right] \mathcal{D}\left[ \bar{\Psi}^\eta,\Psi^\eta \right] e^{i S} \, ,
    \label{GenI}
\end{align}
where we have excluded gauge fixing and Faddeev-Popov fields, and have used fermions as a representative for all matter fields to avoid clutter where it is not necessary. Superscripts $\eta$ and $g$ label the native world of each field. There remains, hence, no ambiguity as to which metric or vielbein should be used for raising and lowering indices of the fields.

An example of a typical action containing intra-world and inter-world bilinear terms is the following,
\begin{equation}
    \tilde{S} = \int d^4x \left[ \bar{\Psi}^g \Psi^g + \bar{\Psi}^g \Psi^\eta +\bar{\Psi}^\eta \Psi^g +\bar{\Psi}^\eta \Psi^\eta \right] \, .
    \label{STyp}
\end{equation}
Variables $\Psi^{g,\eta}$ are defined to be carrying a fourth root of the corresponding metric determinant as $\Psi^g\equiv \sqrt[4]{|g|}\psi^g$ with $\psi^g$ being independent of metric. Writing the action in terms of $\psi^{g,\eta}$ makes the diffeomorphism invariance manifest. The weight half density variables $\Psi^{g,\eta}$ and their consequent measures $\mathcal{D}\Psi^{g,\eta}$ are suitable for curved background considerations (cf. \cite{HawkingZeta,FujikawaBook,FujikawaGRMeasure}). Looking at the above we see that along with the cross-terms in $S_c$ comes the crossbreed volume element $\sqrt[4]{g\eta} d^4x$. Consequently, out of the bi-world construction a new inter-world vacuum energy density will emerge which couples to gravity as $2\bar\Lambda\sqrt[4]{g\eta}$ in the effective form of $S_c$.

Interestingly, $2\bar\Lambda \sqrt[4]{g\eta} $ is the on-shell value of $\bar\Lambda \sqrt{|g|}e^\chi + \bar\Lambda \sqrt{|\eta|}e^{-\chi}$ when solving for the auxiliary scalar $\chi$, which describes the simultaneous concentration and dilution of vacuum energy density in g-world and $\eta$-world respectively. The saddle-point occurs at $e^\chi=\sqrt[4]{\eta/g}$. The ratio of the metric determinants, which are tensor densities of the same weight, is a scalar under general coordinate transformations: $\eta$ transforms in exactly the same way as $g$ does and for their ratio the resulting Jacobians cancel.

Terms that describe a vacuum energy density, in Einstein's equations are proportional to metric, i.e. $\rho g_{\mu\nu}$. Since metric is locally Lorentz invariant, boosting observers does not result in them measuring a different vacuum energy density, therefore they remain unable to find a preferred inertial frame only by looking at a stationary vacuum. Terms in Einstein's equations that are proportional to metric, come from those terms in the action that depend on metric only through its determinant. With just one metric, the only such term we can construct, while preserving diffeomorphism invariance, is the four volume $\int d^4x \sqrt{|g|}$. With two metrics, however, we will have a new scalar, $\sqrt[4]{\eta/g}$, and now we can use any power of this scalar to construct various terms that can describe vacuum as explained above. Let us define $\phi^2 \equiv \sqrt[4]{\eta/g}$ for future convenience.

If either metric, or both, are dynamical we can expect a kinematic term for $\phi$ to appear in the action. Therefore, $\phi$ becomes a suitable candidate for a field that describes the vacuum and possibly its dynamical relaxation. But $\phi$ is more than that. If we factor out a square root of metric determinant from terms in $S_c$ we can write $S_c$ seemingly with a regular volume element. Then, considering the fact that fields hop from one world to another, we can anticipate the following few instances in $S_c$,
\begin{align}
    S_c = \int &d^4x \sqrt{|g|}\Big[ m\phi^2 \bar\psi^g \psi^\eta + w \phi^2 G^\eta_\mu G^{g\mu} +  \nonumber \\ 
    & \alpha \left( \phi^2 + \phi^{-2} \right) \epsilon^{\mu\nu\rho\sigma} F^g_{\mu\nu}F^\eta_{\rho\sigma} 
     + g \leftrightarrow \eta \Big] + \cdots \, ,
     \label{ScPheno}
\end{align}
where $m$, $w$ and $\alpha$ are coupling parameters which symbolically represent inter-world hopping between different fermionic fields, gauge fields and field strengths respectively. Also $\epsilon^{\mu\nu\rho\sigma} \equiv \varepsilon^{\mu\nu\rho\sigma}/\sqrt{|g|}$ is the Levi-Civita tensor with $\varepsilon^{\mu\nu\rho\sigma}=\pm 1$ being the totally anti-symmetric symbol. Therefore, $\phi$ has a Higgs characteristic as well. Other possible terms to include in $S_c$ are variations of $\int d^4x \sqrt{|g|} \phi^2 \bar\psi^\eta \slashed{D}^g \psi^g $ with $\slashed{D}^g$ being the Dirac operator on the curved geometry given by $g_{\mu\nu}$. This of course will modify the chiral anomaly and the non-conservation of the chiral current $\nabla^g_\mu j^\mu_5 = \mathcal{A}_5$  will have additional contributions that are coupled to $\phi$ such as $\phi^2 F\wedge F \subset \mathcal{A}_5$ which describes a scalar boson decay into vector bosons carried through a triangle of fermions \cite{HiggsDecay}. It is also worth noting that the first term in the second line of Eq. \ref{ScPheno} is a naturally arising axionic term \cite{WilczekAxion,WeinbergAxion,MaxwellAxion}, however, a detailed and rigorous investigation of all implications of the moir{\'e} field $\phi$ will be considered elsewhere. Below, we instead focus on its potential cosmological consequences.

\begin{figure}
\includegraphics[width=\linewidth]{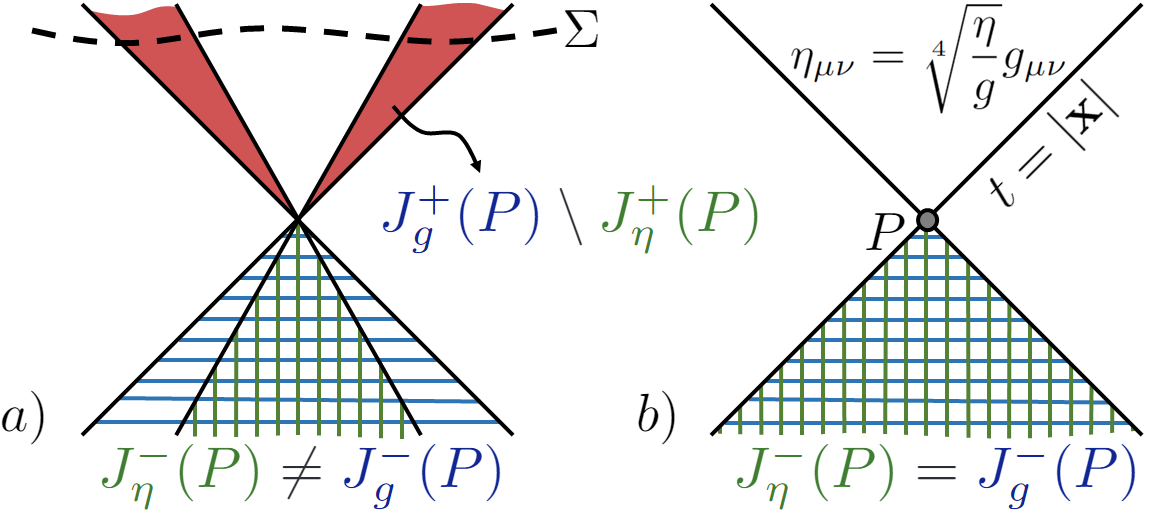}
\centering
\caption{Events that affect $P$, namely its past $J^-(P)$, reside in the hatched regions. Horizontal hatching designates the past of $P$ according to $g_{\mu\nu}$ while vertical hatching is according to $\eta_{\mu\nu}$. (a) The two pasts and futures of $P$ do not match in general. The region $\xi \equiv \Sigma \cap J^+_g(P) \setminus J^+_\eta(P)$ is then effectively forbidden for excitations of $\hat a (x)$. Therefore, in the space of all configurations of $g_{\mu\nu}$ and $\eta_{\mu\nu}$ those with coinciding light cones, as in (b), are statistically favored by an exponential factor of the volume of $\xi$.}
\label{fig:Cones}
\end{figure}

\textit{Bi-causality}---On a manifold with only one metric, say $g_{\mu\nu}$, it is always possible to transform the coordinates to one in which the metric is locally Minkowskian where the line element is expressed as $ds^2_g = -dt^2 + \vec{dx}^2$. This of course is not generally possible for two metrics: We can not always find a local basis in which the line elements match as $ds^2_{g,\eta}=-dt^2 + \vec{dx}^2$. (See Fig. \ref{fig:Cones}) Among other things, this means that in general the light cones of g-world and $\eta$-world do not coincide. On the whole, the gauge classes of a bi-metric $(g_{\mu\nu},\eta_{\mu\nu})$ are smaller than the product of their separate gauge classes.

Metric is responsible both for setting up the causal structure of spacetime and for determining the four-volume element at each point on it. The latter is given by the metric determinant and the former by the remaining degree of freedom in four dimensions. An observer can, in principle, reconstruct the causal structure by gathering which measurements performed at different points of spacetime affect each other. Measurements with a space-like separation must be independent, and hence we expect, the commutators $[a(x),a(y)]$ with $(y-x)^2>0$ to vanish \cite{Peskin}. So microcausality ensures that superluminal propagations are exponentially suppressed; but superluminal with respect to which metric?

Consider, for example, an amphibian field $a(x)$. Using the local commutator data $[a(x),a(y)]$ at a point $P$, segregating disconnected events to $P$ from the rest, we should be able to sketch the light cones of both g-world and $\eta$-world at $P$ simultaneously. This means the two metrics must eventually end up sharing their causal structure. Therefore, one degree of freedom is effectively thrown away and what remains is a causal structure and two metric determinants. Let us look at the probability amplitude of creating $a$ at $P$ and finding it at a later time anywhere,
\begin{align}
    \int_V \langle a(x)a(P) \rangle \equiv \frac{\int_V \int \mathcal{D}g_{\mu\nu}\mathcal{D}\eta_{\mu\nu} \mathcal{D}a \, e^{iS}a(x)a(P)}{\int \mathcal{D}g_{\mu\nu}\mathcal{D}\eta_{\mu\nu} \mathcal{D}a \, e^{iS}} \, .
\end{align}
Since the causal structures described by the two metrics do not necessarily match, $P$ in general has two futures and two pasts, each given by one of the metrics. The path-integral is invariant under coordinate transformations, so we can always locally choose a coordinate where the fastest light cone is on $t=\pm |\vec x|$. The integral $\int_V$ is over all points on an arbitrary space-like hypersurface $\Sigma$ that avoids crossing the past of $P$. But since the propagator is exponentially suppressed for all $x$ that sit outside any of the two light cones, the integral effectively only covers the events inside the narrowest future light cone. Therefore, when we are integrating over all $g_{\mu\nu}$ and $\eta_{\mu\nu}$, those configurations of the two metrics which do not share causal structures have, in effect, less accessible volume and hence exponentially less microstates in comparison to those with coinciding light cones. The light cones are sketched by $ds^2_{g,\eta}=0$. Thus, when the causal structure is set, there remains a conformal degree of freedom, as in $\phi^2 ds^2_g = \phi^2 (-dt^2 + \vec{dx}^2)= ds^2_\eta$ where the light cones of both worlds are coinciding on $t=\pm |\vec x|$. Consequently, the contributing configurations are expected to approximately be those with conformally related metrics $$
\eta_{\mu\nu}=\sqrt[4]{\frac{\eta}{g}}\, g_{\mu\nu} = \phi^2 g_{\mu\nu},
$$
which directs us to rewrite the geometrical part of  path-integration \eqref{GenI} in terms of the moir{\'e} field, $\phi$, as follows
\begin{equation}
    \int \mathcal{D}g_{\mu\nu} \mathcal{D}\eta_{\mu\nu} e^{iS[g_{\mu\nu},\eta_{\mu\nu}]} \approx \int \mathcal{D}g_{\mu\nu} \mathcal{D}\phi\,  e^{iS[g_{\mu\nu},\phi^2 g_{\mu\nu}]} .
    \label{PIChange}
\end{equation}


\textit{Inflation and dynamic relaxation}--- Let us now, for the sake of simplicity, resort to considering only two dynamical metrics with the usual Einstein-Hilbert action each, and minimize the appearance of matter fields to three accumulated vacuum energy densities,
\begin{align}
    S = \int d^4 x \Big[ &\sqrt{|g|} \left( R^g - 2\Lambda^g \right)  \label{SEH} \\
    &+ \sqrt{|\eta|} \left( R^\eta - 2\Lambda^\eta \right) - 4\bar\Lambda \sqrt[4]{g\eta} \Big] \, , \nonumber
\end{align}
where $R^{g,\eta}$ are the Ricci scalars built out of each metric, $\Lambda^{g,\eta}$ the intra-world cosmological constants and $\bar\Lambda$ the inter-world cosmological constant coming from the tunneling terms. Classically, it is enough to solve the Einstein's equations derived from taking the variation of the above action. $\delta S/\delta g^{\mu\nu}$ yields,
\begin{equation}
    R^g_{\mu\nu} - \frac{1}{2} R^g g_{\mu\nu} + \left( \Lambda^g + \bar\Lambda \sqrt[4]{\frac{\eta}{g}} \right) g_{\mu\nu} =0 \, ,
    \label{ClassicalEOM}
\end{equation}
and the equations of motion for $\eta_{\mu\nu}$ are given by $g\leftrightarrow \eta$. With $\Lambda^g$ and $\bar\Lambda$ being constants, the contracted Bianchi identity makes sure that $\phi^2 = \sqrt[4]{\eta/g}$ is constant as well. This constant will be dictated by the equations of motion and the symmetries imposed on the desired solution. For instance, for the case of $\Lambda^g = \Lambda^\eta$ any value of $\phi^2$ other than one will result in different vacuum energy densities for $g_{\mu\nu}$ and $\eta_{\mu\nu}$ which for an expanding universe means a time varying ratio of metric determinants, in contradiction with a changeless $\phi$. Thus in that case $\phi^2=1$ classically. However, we will see that the quantum effects discussed in the previous section will allow $\phi$ to vary. But we expect such quantum gravitational deviations from the classical theory only at huge energy or temperature scales. Therefore, the effective theory should  descend to the classical regime in the low-energy limit (which is equivalent to the long-time limit in the context of cosmology).

Following \eqref{PIChange} we can change the action \eqref{SEH} to its effective form (see Supplemental Material for the required conformal transformations) by substituting $\eta_{\mu\nu} \rightarrow \phi^2 g_{\mu\nu}$ everywhere,
\begin{align}
    S = \int d^4 x \sqrt{|g|} \Big[ &\left(1 + \phi^2 \right) R + 6 k \partial_\mu \phi \partial^\mu  \phi \label{SPhi} \\
    & \quad -2 \Lambda^\eta \phi^4 -  2\Lambda^g - 4 \bar\Lambda \phi^2 \Big] \, . \nonumber
\end{align}
Here we have omitted the surface terms and $k$ represents the effect of unstated parts of the action \eqref{SEH} on the kinetic term for $\phi$. We have also dropped the superscript on $R\equiv R^g$. The above action has one metric $g_{\mu\nu}$ and one scalar $\phi^2$ as its dynamical degrees of freedom which are going to be integrated over in the path-integral approach. Therefore, apart from its similarities to Higgs and a vacuum order parameter, $\phi^2$ also resembles a dilaton field.

The action \eqref{SEH}, and consequently the above action \eqref{SPhi}, are devised for Neumann boundary conditions \cite{Krishnan}. This fact particularly means that in the effective action the initial condition for $\phi$ is set by fixing its derivative on a generic Cauchy surface, so that we can begin in an ``slow-roll'' regime where we can disregard the kinetic term. Then the saddle-point approximation, using the on-shell value of $\phi^2 \doteq (R-4\bar\Lambda)/4\Lambda^\eta$, will give us a renormalized Starobinsky action: 
\begin{equation}
    S \doteq \int d^4x \sqrt{|g|} \left[ R - 2\Lambda^g + \frac{\left(R - 4\bar\Lambda\right)^2}{8\Lambda^\eta} \right] \, .
\end{equation}
The Starobinsky action \cite{Starobinsky,StarobinskyAction} and in general higher curvature terms can arise from quantum gravity considerations \cite{DaviesPL}. The recurrence of the above, therefore, might be a hint towards the bi-world construction having a similar origin.

To achieve a minimally coupled theory of gravity out of \eqref{SPhi} we conformally transform our (now only) metric by $g_{\mu\nu} \rightarrow g_{\mu\nu}/(1+\phi^2)$ to get,
\begin{align}
    S = \int d^4x \sqrt{|g|} \Bigg[ R &- 6\frac{(1-k)\phi^2-k}{(1+\phi^2)^2} \partial_\mu \phi \partial^\mu \phi \nonumber \\ 
     & \ \quad -2 \frac{\Lambda^\eta \phi^4 + \Lambda^g + 2\Bar\Lambda \phi^2}{(1+\phi^2)^2} \Bigg] \, ,
     \label{SDecoupled}
\end{align}
where the dynamics governing $\phi$ is more clear. Since we are interested in the cosmological consequences here we will simplify the model by choosing the Friedmann-Lemaître–Robertson–Walker metric $ds^2 \equiv g_{\mu\nu}dx^\mu dx^\nu = -dt^2 + a^2(t)d\vec{x}^2$ as an ansatz to the equations of motion. These are,
\begin{align}
    \ddot{\phi} = -3H\dot\phi +\frac{\phi}{(1-k)\phi^2 - k} \Bigg[ \frac{(1-k)\phi^2-k-1}{1+\phi^2}\dot{\phi}^2 \nonumber \\ - \frac{2}{3}\frac{( \Lambda^\eta - \bar\Lambda )\phi^2 - (\Lambda^g - \bar\Lambda)}{1+\phi^2} \Bigg] \, ,
    \label{Phiddot}
\end{align}
with $H \equiv \dot{a}/a$ being the Hubble parameter. Depending only on time, $\phi (t)$ now has the dynamics of a single particle rolling on a potential. The first and second terms on the right hand side are friction exerted on this particle while the last one is the conservative force which is always zero at $\phi^2 = (\Lambda^\eta - \Bar\Lambda)/(\Lambda^g - \bar\Lambda)$. The potential corresponding to this force is a Mexican hat except for $k=0$ where the crown of the hat is uplifted to infinity. The constraint part of Einstein's equations is given by,
\begin{equation}
	3H^2 = 3\frac{(1-k)\phi^2 - k}{(1+\phi^2)^2} \dot{\phi}^2 + \frac{\Lambda^\eta \phi^4 + \Lambda^g + 2\bar\Lambda \phi^2}{(1+\phi^2)^2} \, ,
	\label{Constraint}
\end{equation}
while the dynamical part is,
\begin{equation}
	2\dot{H}+3H^2 = -3\frac{(1-k)\phi^2 - k}{(1+\phi^2)^2} \dot{\phi}^2 + \frac{\Lambda^\eta \phi^4 + \Lambda^g + 2\bar\Lambda \phi^2}{(1+\phi^2)^2} \, ,
\end{equation}
which combine to simply give,
\begin{equation}
	\dot{H} = -3\frac{(1-k)\phi^2 - k}{(1+\phi^2)^2} \dot{\phi}^2 \, .
	\label{Hdot}
\end{equation}
See Supplemental Material for a derivation of the action functional \eqref{SDecoupled} and the equations of motion above.

\begin{figure}
\includegraphics[width=\linewidth]{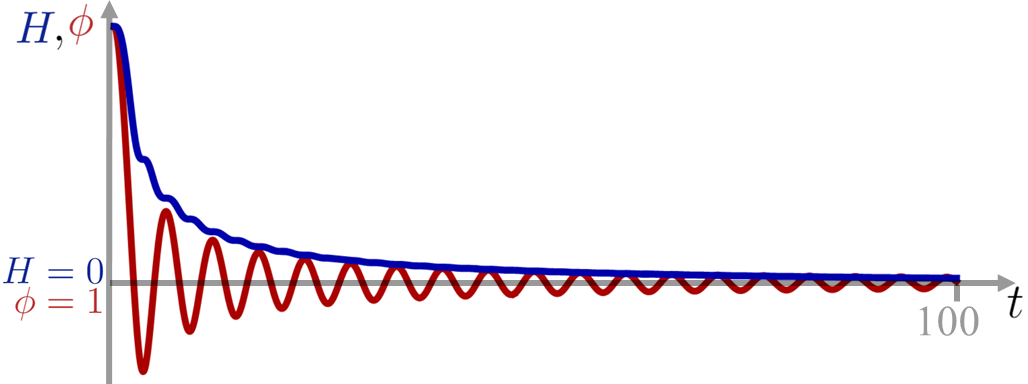}
\centering
\caption{Evolutions of the moir\'e field $\phi$ and the Hubble parameter $H$ in units of Planck time $t_P$. Initially the worlds are away from equilibrium by $\phi=1.5$. Here $k$ is set to zero while the cosmological constants are set to be $\Lambda^{g,\eta}=-\bar\Lambda=M_P^2$. $H$ (the plot on top in blue) starts from $0.22 M_P$ and approaches \underline{zero} as time goes by while $\phi$ starts from $1.5$ and oscillates around $\phi=\underline{1}$ henceforward. Whenever the rate of change in $\phi$ is small the universe experiences a period of stationary $H$. }
\label{fig:HPhi}

\end{figure}

Although not necessary for our purposes, hereon we adopt the symmetric situation where the two worlds are similar whence the desired destiny is for them to eventually reach equilibrium at $\phi^2 = 1$. According to \eqref{Hdot}, for all values of $k$ below $1/2$ the Hubble parameter will be ever decreasing up to the point of equilibrium. Therefore, we can imagine the story of the universe beginning with the two worlds either already out of equilibrium or kicked out of it, while according to equation \eqref{Constraint} the state of non-equilibrium, $\phi^2 \neq 1$, is accompanied with a huge vacuum energy density. All this however will be relaxed when the worlds eventually reach equilibrium with the help of the friction forces. An example has been plotted in Fig. \ref{fig:HPhi}. At equilibrium $\phi$ remains constant and hence the classical theory \eqref{ClassicalEOM} is reinstated.

\begin{figure}
\includegraphics[width=\linewidth]{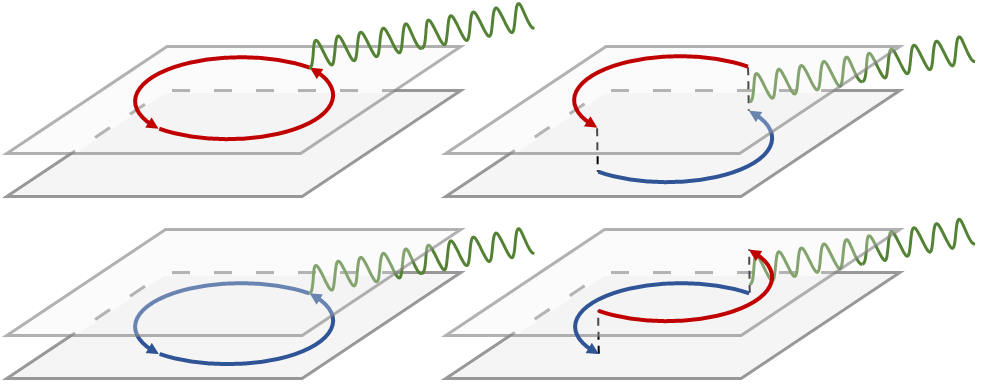}
\centering
\caption{Coupling of external gravitons (wiggly lines) to vacuum loops (solid lines) of intra-world vacua (on the left) and inter-world vacua (on the right). }
\label{fig:VacGrav}
\end{figure}

Having two similar worlds implies that their characteristic parameters (coupling constants, cosmological constants, etc.) are the same, e.g. $\Lambda^\eta = \Lambda^g \approx M_P^2$. This resembles a system of two overlapping grids with characteristic lengths of $L_P$. The zero of the conservative force in \eqref{Phiddot} happens at $\phi^2=1$. As discussed before, the classical limit further requires this point of equilibrium to be stable which is met only if $\bar\Lambda$ has a different sign than $\Lambda^{g,\eta}$. On the other hand, the cross terms in $S_c$ are built out of the same fields and parameters appearing in $S^{g,\eta}$. There is an inter-world hopping term from g-world to $\eta$-world in addition to its conjugate which describes the reverse hopping. Therefore the magnitude of inter-world vacuum energy density must be twice that of the intra-world ones in the action \eqref{SEH}. We can reinforce this picture by looking at a few possible diagrams of the gravitating bi-world vacuum Fig. \ref{fig:VacGrav}
where for each vacuum loop in a single world there is a counterpart that passes through both worlds. All in all, we have $\Lambda^{g,\eta}=-\bar\Lambda \approx M_P^2$. This combination yields the desired history for the universe (Fig. \ref{fig:HPhi}) starting from an inflationary period and reaching a small vacuum energy density that is continuously approaching zero at late times.


\textit{Conclusion}---Although the classical limit in a way ensures it, a robust and detailed consideration of why the cosmological constants have such a relation requires a solid knowledge of the origin of the bi-world in the proposed scenario. Discounting the classical limit we are still left with some degree of maneuver. For example, we know that the inter-world quantum vacuum is constructed out of the same fields and processes as the intra-world quantum vacuum. Since there are no new scales introduced in the theory we expect the vacuum energy densities to be of equal magnitude, which leaves us merely with deciding what sign must be attributed to inter-world vacuum energy density. It is worth noting that $\mathbb{Z}^{g,\eta}_2$ invariant terms in $S_{g,\eta}$ have counterparts in $S_c$ which do not remain invariant under either $\mathbb{Z}^g_2$ or $\mathbb{Z}^\eta_2$ separately. Thus one might naively decide that the sign choice is arbitrary and/or a property attributed to a symmetry broken universe. In any case, the negative sign fits our picture more naturally: If the worlds were lying on membranes separated by some distance as in Fig. \ref{fig:VacGrav} the tunneling terms would have exerted an attractive or repulsive force between them and their distance would have settled on the point of equilibrium between these forces. Albeit the separation in Fig. \ref{fig:VacGrav} is only for illustration purposes. There is no separation between the membranes in the bi-world structure which in membrane picture translates to an always attractive force between the membranes. This in turn means that the corresponding potential for the inter-layer force appears in the Lagrangian as $\mathcal{L} \subset -V(l)$ with $l$ being the separation between the membranes and $\dd {V}{l}>0$. For terms in $S_c$ to result in such a potential they must appear with a minus sign themselves which results in a negative inter-world vacuum energy density given the intra-world ones are positive.

The theory presented here is experimentally falsifiable also through considering the detailed behaviour of the moir\'e field $\phi$. For example, the axion term in \eqref{ScPheno} when integrated out will modulate the effective moir\'e field potential in an oscillatory manner \cite{PecceiQuinn,Abbott} near the extrema, which possibly leaves a trace to be found by early universe physics observations.

Lastly, it should be mentioned that the construction briefly sketched in this Letter differs from theories of bimetric or massive gravity \cite{Bimetric,BimetricNon,BimetricMass} in several respects of which we will emphasise on the following: The metrics couple to each other and to matter in a different manner as we have here; the current construction generally describes a universe containing two worlds rather than only two metrics, it includes matter fields the effects of which are crucial and measurable at least through cosmological observations; there is an emergent inherently geometric moir\'e field $\phi$ that plays multiple roles within the theory; gravity remains massless here; and the foundational philosophy is different promising a feasible origin story. (For a more detailed discussion see the Supplemental Material.)



\acknowledgements This work originated from a project supported by the Templeton Foundation. This research was also supported by the Simons Foundation. A.P. wishes to thank Batoul Banihashemi and Ted Jacobson for useful discussions. 

\bibliographystyle{apsrev4-1}
\bibliography{main}


\end{document}


\title{
		Moir\'e Gravity and Cosmology
		\\
		Supplemental Material
	}
	\author{Alireza Parhizkar}
	\author{Victor Galitski}
	\affiliation{Joint Quantum Institute and Condensed Matter Theory Center, Department of Physics, University of Maryland, College Park, MD 20742, USA.}
	
	\date{\today}

	\maketitle
	
	\tableofcontents

\section{Conformal Transformations}
A conformal transformation that takes $g_{\mu\nu}$ to $\tilde g_{\mu\nu} \equiv \Omega^2 g_{\mu\nu}$, also takes $g^{\mu\nu}$ to $\tilde g^{\mu\nu} = \Omega^{-2} g^{\mu\nu}$, and $g$ to $\tilde g = \Omega^8 g$ in a four-dimensional space-time. The chain rule divides the transformed Christoffel symbols $\tilde \Gamma^\alpha_{\mu\nu}=\frac{1}{2}\tilde g^{\alpha\beta}\left(-\partial_\beta \tilde g_{\mu\nu} + \partial_\mu \tilde g_{\beta\nu} + \partial_\nu \tilde g_{\beta\mu} \right)$ into the untransformed one plus derivatives of $\Omega$, so that we have,
\begin{align}
    \tilde{\Box} \phi \equiv \tilde g^{\mu\nu} \tilde \nabla_\mu \partial_\nu \phi &= \Omega^{-2}g^{\mu\nu} \left[\partial_\mu\partial_\nu \phi - \tilde\Gamma^\alpha_{\mu\nu} \partial_\alpha \phi \right] \\
    &= \Omega^{-2}g^{\mu\nu} \left[ \nabla_\mu\partial_\nu \phi - \Omega^{-1}\left(  \partial_\mu\Omega \partial_\nu\phi + \partial_\nu\Omega \partial_\mu\phi - g_{\mu\nu}g^{\alpha\beta}\partial_\beta\Omega\partial_\alpha\phi \right) \right] \\
    &= \Omega^{-2}\left[\Box \phi + 2 \Omega^{-1} g^{\mu\nu}\partial_\mu\Omega\partial_\nu\phi \right] \, .
\end{align}

We use the $(-+++)$ metric signature throughout the Letter and the supplemental material, in addition to using the following conventions:
\begin{align}
R^\alpha_{\ \mu \beta \nu} &= \partial_\beta \Gamma^\alpha_{\mu\nu} - \partial_\nu\Gamma^\alpha_{\mu\beta} + \Gamma^\alpha_{\beta\sigma}\Gamma^\sigma_{\mu\nu} - \Gamma^\alpha_{\nu\sigma}\Gamma^\sigma_{\beta\mu} \, , \\
R_{\mu\nu} &= R^\alpha_{\mu\alpha\nu} \, .
\end{align}
With the above conventions the Ricci tensor and scalar transform as,
\begin{align}
    \tilde R_{\mu\nu} &= R_{\mu\nu} + \Omega^{-2}\left( 4 \partial_\mu \Omega \partial_\nu \Omega - 2\Omega\nabla_\mu\partial_\nu \Omega - g^{\alpha\beta}\partial_\alpha \Omega \partial_\beta \Omega \, g_{\mu\nu} - \Omega \Box \Omega \, g_{\mu\nu} \right) \, , \\
    \tilde R &= \Omega^{-2} R-6\Omega^{-3}\Box\Omega \, ,
\end{align}
and transform reversely as,
\begin{align}
     R_{\mu\nu} &= \tilde R_{\mu\nu} + \Omega^{-2} \left(  2 \Omega \tilde\nabla_\mu\partial_\nu\Omega - 3 \tilde g^{\alpha\beta} \partial_\alpha \Omega \partial_\beta \Omega \, \tilde g_{\mu\nu} + \Omega \tilde\Box \Omega \, \tilde g_{\mu\nu} \right) \, , \\
      R &= \Omega^2 \tilde R + 6 \Omega\tilde\Box\Omega - 12 \tilde g^{\mu\nu}\partial_\mu\Omega\partial_\nu\Omega \, .
\end{align}
The above constitute all the conformal transformations we need for the next section.


\section{Starobinsky Action}

Starting from the bi-world action, substituting $\eta_{\mu\nu}$ with $\phi^2 g_{\mu\nu}$ which means $\eta \rightarrow \phi^8 g$ and $R[\eta_{\mu\nu}] \rightarrow R[\phi^2 g_{\mu\nu}]$, and employing the relations given in the previous section, we will have,
\begin{equation}
    S = \int d^4x \sqrt{|g|} \left[ \left(1+\phi^2\right) R + 6 k g^{\mu\nu} \partial_\mu \phi \partial_\nu \phi - 2\Lambda^\eta \phi^4 - 2\Lambda^g - 4\bar{\Lambda} \phi^2 \right] \, ,
    \label{supCoupled}
\end{equation}
where we have summarized the effect of additional kinetic terms coming from the effective theory in the coefficient $k$. Looking at the non-minimally coupled action above, we can imagine a regime in which $\phi$ is slow-rolling and its kinetic term is negligible, so that we roughly have,
\begin{equation}
    S = \int d^4x \sqrt{|g|} \left[ \left(1+\phi^2\right) R - 2\Lambda^\eta \phi^4 - 2\Lambda^g - 4\bar{\Lambda} \phi^2 \right] \, .
    \label{supSlow}
\end{equation}
Solving for $\phi$ in this regime now gives at the saddle point,
\begin{equation}
    2\phi R - 8\Lambda^\eta \phi^3 - 8\bar\Lambda \phi = 0 \, ,
\end{equation}
or
\begin{equation}
    \Phi \equiv \phi^2 = \frac{R-4\bar\Lambda}{4\Lambda^\eta}  \, .
\end{equation}
Substituting the saddle-point value above back to the action \eqref{supSlow} we get,
\begin{align}
    S &= \int d^4x \sqrt{|g|} \left[R - 2\Lambda^g  + (R - 4\bar\Lambda) \phi^2 - 2\Lambda^\eta \phi^4 \right] \nonumber \\
    &= \int d^4x \sqrt{|g|} \left[ R- 2\Lambda^g + \frac{(R-4\bar\Lambda)^2}{4\Lambda^\eta}  - \frac{(R-4\bar\Lambda)^2}{8\Lambda^\eta}\right] \nonumber \\
    &= \int d^4x \sqrt{|g|} \left[ R - 2\Lambda^g + \frac{(R-4\bar\Lambda)^2}{8\Lambda^\eta} \right]
    \, .
\end{align}
We can arrive at the same action also by squaring the $\Phi$ field out:
\begin{align}
    S &= \int d^4x \sqrt{|g|} \left[ \left(1+\Phi\right) R - 2\Lambda^\eta \Phi^2 - 2\Lambda^g - 4\bar{\Lambda} \Phi \right] \nonumber \\
    &= \int d^4x \sqrt{|g|} \left[ R - 2\Lambda^g + \Phi (R - 4\bar\Lambda) - 2\Lambda^\eta \Phi^2 \right] \nonumber \\
    &= \int d^4x \sqrt{|g|} \left[ R - 2\Lambda^g + \frac{(R-4\bar\Lambda)^2}{8\Lambda^\eta} -2\Lambda^\eta \left( \Phi - \frac{R-4\bar\Lambda}{4\Lambda^\eta}  \right)^2 \right]
    \, .
\end{align}




\section{Transformation to Einstein Frame}

Let us go back to the action \eqref{supCoupled} introduced at the beginning of the previous section:
\begin{equation}
    S = \int d^4x \sqrt{|g|} \left[ \left(1+\phi^2\right) R + 6 k g^{\mu\nu} \partial_\mu \phi \partial_\nu \phi - 2\Lambda^\eta \phi^4 - 2\Lambda^g - 4\bar{\Lambda} \phi^2 \right] \, , \nonumber
\end{equation}
where the Ricci scalar is coupled to $\phi$ through $\int d^4x \sqrt{|g|}\left(1+\phi^2\right)R$.
Hoping to decouple $\phi$ from $R$, we are going to choose, $\tilde{g}_{\mu\nu} = (1+\phi^2) g_{\mu\nu}$ and therefore have $\sqrt{|\tilde g|} =(1+\phi^2)^2 \sqrt{|g|}$, and write the action above in terms of $\tilde{g}_{\mu\nu}$ where everything with tilde sign is built out of this metric. By only transforming the first term in the Lagrangian we will get,
 \begin{align}
     S=\int d^4x \sqrt{|\tilde{g}|} (1+\phi^2)^{-2} \Bigg[ (1+\phi^2)^2 \left( \tilde{R} + \frac{6}{\sqrt{1+\phi^2}}\tilde{\Box}\sqrt{1+\phi^2} - \frac{12}{1+\phi^2}\tilde{g}^{\mu\nu} \partial_\mu \sqrt{1+\phi^2}\partial_\nu\sqrt{1+\phi^2}\right) \nonumber \\
     + 6 k g^{\mu\nu} \partial_\mu \phi \partial_\nu \phi - 2\Lambda^\eta \phi^4 - 2\Lambda^g - 4\bar{\Lambda} \phi^2 \Bigg] \, .
 \end{align}
We can see why this transformation is helpful for decoupling $\phi$ from the curvature scalar. Let us set $\Omega \equiv \sqrt{1+\phi^2}$ to avoid further clutter so that $g^{\mu\nu} = \Omega^2 \tilde{g}^{\mu\nu}$. Now notice that the d'Alembertian term $6\tilde{\Box} \Omega / \Omega$ divides into a boundary term and $6\partial_\mu \Omega \partial^\mu \Omega / \Omega^2$.
\begin{align}
    S &= \int d^4x \sqrt{|\tilde{g}|} \Bigg[ \tilde{R} -\frac{6}{\Omega^2}\tilde{g}^{\mu\nu} \partial_\mu\Omega \partial_\nu\Omega + \Omega^{-4} \left( 6k\frac{1+\phi^2}{\phi^2}g^{\mu\nu}\partial_\mu\sqrt{1+\phi^2}\partial_\nu\sqrt{1+\phi^2} - 2\Lambda^\eta\phi^4 - 2\Lambda^g - 4\bar\Lambda \phi^2 \right) \Bigg] \\
    &= \int d^4x\sqrt{|\tilde g|} \Bigg[ \tilde{R} -\frac{6}{\Omega^2}\tilde{g}^{\mu\nu} \partial_\mu\Omega \partial_\nu\Omega + \frac{6k}{\Omega^2 -1} \tilde{g}^{\mu\nu}\partial_\mu\Omega\partial_\nu\Omega - 2\Lambda^\eta \frac{(\Omega^2-1)^4}{\Omega^{4}} - 2\Lambda^g \frac{1}{\Omega^4} - 4\bar\Lambda \frac{\Omega^2-1}{\Omega^4}  \Bigg] \, .
\end{align}
The factor behind the kinetic term is proportional to $-1/\Omega^2 + k/(\Omega^2 - 1)$ which for a stable evolution we will want to be negative. In terms of $\phi$ this condition is translated to $\phi^2>k/(1-k)$. If we require $\phi$ to be able to reach the equilibrium value $\phi=1$ of the symmetric case, then the lower limit of $\phi$ is one. Which determines the allowed values of $k$ to be $-\infty<k<1/2$. Let us remove the tilde sign and use $\phi$ again which was our original physical degree of freedom to eventually get,
\begin{equation}
    S = \int d^4x \sqrt{|g|} \Bigg[ R - 6\frac{(1-k)\phi^2 - k}{(1+\phi^2)^2} \partial_\mu \phi \partial^\mu \phi -\frac{2}{(1+\phi^2)^2}\left( \Lambda^\eta \phi^4 + \Lambda^g + 2\bar\Lambda\phi^2\right)  \Bigg] \, .
    \label{SDecoupled}
\end{equation}


\section{Equations of Motion}

Now that $\phi$ is decoupled from the curvature scalar, it is easier to derive the equations of motion. By varying the action \eqref{SDecoupled} with respect to $\phi$ while other degrees of freedom are kept fixed we will have,
\begin{align} \label{PhiEOM1}
   \delta S \Big|_{g^{\mu\nu}} &= \int d^4x \sqrt{|g|} \Bigg[ -12\phi\frac{1+k-(1-k)\phi^2}{(1+\phi^2)^3}\partial_\mu \phi \partial^\mu\phi \delta \phi + 6\frac{k-(1-k)\phi^2}{(1+\phi^2)^2} g^{\mu\nu} \nabla_\mu \phi \nabla_\nu \delta\phi \nonumber \\
   &\quad\quad\quad\quad\quad\quad\quad - 8 \Lambda^\eta \left( \frac{\phi^2}{1+\phi^2} \right) \frac{\phi\delta\phi}{(1+\phi^2)^2} + 8 \Lambda^g \frac{\phi\delta\phi}{(1+\phi^2)^3} - 8\bar{\Lambda}\left(\frac{1-\phi^2}{1+\phi^2} \right) \frac{\phi\delta\phi}{(1+\phi^2)^2} \Bigg] \\
   &= \int d^4x \sqrt{|g|} \Bigg[ 12\phi\frac{1+k-(1-k)\phi^2}{(1+\phi^2)^3}\partial_\mu \phi \partial^\mu\phi \delta \phi + 12\frac{(1-k)\phi^2 -k}{(1+\phi^2)^2}\Box\phi \delta\phi \nonumber \\ &\quad\quad\quad\quad\quad\quad\quad  -8\frac{\phi}{(1+\phi^2)^3} \left(\Lambda^\eta  \phi^2   -  \Lambda^g  + \bar{\Lambda}\left(1-\phi^2 \right)  \right)\delta\phi \Bigg] \, ,
\end{align}
where for the second equality we also applied an integration by parts on the second term in \eqref{PhiEOM1}:
\begin{equation}
    6\frac{k-(1-k)\phi^2}{(1+\phi^2)^2} g^{\mu\nu} \nabla_\mu \phi \nabla_\nu \delta\phi \rightarrow 2\times 12\phi\frac{1+k-(1-k)\phi^2}{(1+\phi^2)^3}\partial_\mu \phi \partial^\mu\phi \delta \phi \, + \,  12\frac{(1-k)\phi^2 -k}{(1+\phi^2)^2}\Box\phi \delta\phi \, + \, \text{Boundary term.}
\end{equation}
Therefore, by requiring $\delta S /\delta \phi = 0$ we arrive at the equations of motion for $\phi$ below,
\begin{equation}
    \phi\frac{1+k -(1-k)\phi^2}{1+\phi^2}\partial_\mu\phi\partial^\mu\phi + ((1-k)\phi^2 - k)\Box \phi - \frac{2}{3}\frac{\phi}{1+\phi^2}\left((\Lambda^\eta-\bar{\Lambda})\phi^2 + \bar{\Lambda} - \Lambda^g \right) =0 \, .
    \label{PhiEq}
\end{equation}

Next we are going to derive the equations motion for the inverse metric $g^{\mu\nu}$. This is however very easy to calculate in the Einstein frame given by action \eqref{SDecoupled}. Since the Ricci scalar is decoupled from $\phi$, the variation of the first term in the Lagrangian, namely $\sqrt{|g|}R$, simply gives the Einstein tensor times $\sqrt{|g|}$. The last term in the Lagrangian is coupled to metric only through square root of the metric determinant and its variation is easily obtained by remembering that $\delta \sqrt{g} = -\frac{1}{2}\sqrt{g} g_{\mu\nu} \delta g^{\mu\nu}$. The kinetic term $\sqrt{|g|}f(\phi)g^{\mu\nu}\partial_\mu\phi\partial_\nu\phi$ is coupled to the metric both by the metric determinant, which just like the last term yields an energy density contribution, and also through $g^{\mu\nu}$ which gives the rank two tensor $\sqrt{|g|}f(\phi)\partial_\mu\phi\partial_\nu\phi$. All in all we have,
\begin{align}
    R_{\mu\nu} - \frac{1}{2}R g_{\mu\nu} + 6\frac{k-(1-k)\phi^2}{(1+\phi^2)^2}\partial_\mu\phi\partial_\nu\phi + \frac{1}{(1+\phi^2)^2}\left(3[(1-k)\phi^2-k]\partial_\mu\phi\partial^\mu\phi + \Lambda^\eta\phi^4 + \Lambda^g + 2\bar\Lambda\phi^2 \right)g_{\mu\nu}=0 \, .
    \label{EinstEq}
\end{align}

We now can simplify the problem by reducing the set of equations to those that enjoy a large class symmetry, namely, the isometric homogeneous solutions. For this purpose we are going to use the Friedmann-Lemaitre-Robertson-Walker (FLRW) metric by setting $g_{\mu\nu}=\text{diag}\left(-1,a^2,a^2,a^2\right)$ with $a\equiv a(t)$ being a function of only $t$. The non-zero Christoffel symbols are,
\begin{align}
    &\Gamma^x_{xt} = \frac{1}{2}g^{xx}\left(\partial_x g_{tx} +\partial_t g_{xx} \right) = \frac{\dot{a}}{a} \equiv H \, , \\
    &\Gamma^t_{xx} = \frac{1}{2}g^{tt}\left(2\partial_x g_{xt}-\partial_t g_{xx} \right) = a\dot{a} = Ha^2 \, .
\end{align}
where $x$ represents all the spatial components. We are also going to use the ansatz $\phi(t,x) = \phi(t)$ for the moir\'e field. The Einstein's equations \eqref{EinstEq} for $\mu=\nu=0$ therefore give,
\begin{align}
    & 3H^2 + 6\frac{k-(1-k)\phi^2}{(1+\phi^2)^2}\dot{\phi}^2 - \frac{1}{(1+\phi^2)^2}\left(3[k - (1-k)\phi^2 ]\dot{\phi}^2 + \Lambda^\eta\phi^4 + \Lambda^g + 2\bar\Lambda\phi^2 \right)=0 \Rightarrow \nonumber \\
    & 3H^2 - \frac{1}{(1+\phi^2)^2}\left(-3[k - (1-k)\phi^2 ]\dot{\phi}^2 + \Lambda^\eta\phi^4 + \Lambda^g + 2\bar\Lambda\phi^2 \right)=0 \, ,
\end{align}
while for $\mu=\nu=x$ they are,
\begin{equation}
    -2\dot{H}^2 - 3H^2 + \frac{1}{(1+\phi^2)^2}\left(3[k-(1-k)\phi^2]\dot{\phi}^2 + \Lambda^\eta\phi^4 + \Lambda^g + 2\bar\Lambda\phi^2 \right) = 0 \, ,
\end{equation}
and they combine to the following simplified form,
\begin{equation}
	\dot{H} = -3\frac{(1-k)\phi^2 - k}{(1+\phi^2)^2} \dot{\phi}^2 \, .
	\label{Hdot}
\end{equation}

We also note that the equation of motion \eqref{PhiEq} for $\phi = \phi(t)$ in the flat FLRW metric reduces to,
\begin{align}
    \ddot{\phi} = -3H\dot\phi +\frac{\phi}{(1-k)\phi^2 - k} \Bigg[ \frac{(1-k)\phi^2-k-1}{1+\phi^2}\dot{\phi}^2 - \frac{2}{3}\frac{( \Lambda^\eta - \bar\Lambda )\phi^2 - (\Lambda^g - \bar\Lambda)}{1+\phi^2} \Bigg] \, ,
    \label{supPhiddot}
\end{align}
where we have noted that,
\begin{equation}
    \Box \phi = g^{\mu\nu}\nabla_\mu\partial_\nu \phi = -\ddot{\phi} - g^{\mu\nu}\Gamma^t_{\mu\nu}\partial_t \phi = -\ddot{\phi} - 3 H \dot{\phi} \, .
\end{equation}


\section{Difference to Bimetric and Massive Theories}

The work presented in the Letter is different from bimetric theories in multiple important respects which are going to be pointed out below.
\begin{itemize}
    \item The bimetric theories have generally been of interest due to their potential in giving rise to massive, and consequently short-range, gravitational modes. This is not the goal of the current work.
    
    Fierz and Pauli \cite{FierzPauli} introduced the unique mass term that leaves the classical linearized theory of gravity, around the flat spacetime, consistent. The tracelessness of the perturbation $h_{\mu\nu}$ around the Minkowski spacetime, is an on-shell constraint that removes an unphysical degree of freedom. Fierz-Pauli mass term preserves this on-shell constraint which otherwise would have given rise to a ghost-like scalar mode. Ghosts are degrees of freedom with negative kinetic energy that lead to instability classically and to non-unitarity at the quantum mechanically~\cite{HawkingGhost}. With the ghost degree out of the way, the Fierz-Pauli model describes a massive spin-2 propagation with five degrees of freedom. Note that the massless gravity comes only with two degrees of freedom.
    
    There are indeed certain issues with a generic massive theory of gravity. One prominent example is the Boulware-Deser ghost \cite{BolDes}. To promote the massive gravity to the non-linear level, one can build up on the model of Fierz and Pauli in order to arrive at its non-linear extension. Boulware and Deser showed that for a wide class of such extensions ghost-like scalar modes are inevitable.
    
    Our model does not use the non-linear extension of the Fierz-Pauli mass \cite{FierzPauli} and is not subject to such ghost problems. To show the persistence of such difficulties the literature usually considers those theories which, in the free, limit reduce to the Fierz-Pauli action; that is irrelevant to the model presented in the Letter where gravity remains massless.
    
    \item In addition to having two metrics we also have two copies of each matter field and a newly introduced amphibian field. Note that the effective theory point of view is that all these terms should be present if they leave the theory self-consistent e.g. the required symmetries are preserved both at the classical and the quantum level.
    
    \item The natural form the inter-world coupling terms are dictated by using the suitable choice of path-integral measures on a curved background. Following the work by Hawking \cite{HawkingZeta} on path-integral measures for quantum field theories on curved spacetimes, we use a specific measure for matter fields. These carry a fourth-root of their corresponding metric determinant and \textit{inevitably} give rise to $\sqrt[4]{g\eta}$ terms, which consequently result in the inter-world cosmological constant. Therefore, the ``bimetric'' coupling is not arbitrarily chosen here, rather it is given through demanding a consistent path-integral formulation.
    
    \item At the classical level the model reduces to regular gravity theories. This can be observed by recalling that the classical equations of motion in our model read:
    \begin{equation}
        R^g_{\mu\nu} - \frac{1}{2} R^g g_{\mu\nu} + \left( \Lambda^g + \bar\Lambda \sqrt[4]{\frac{\eta}{g}} \right) g_{\mu\nu} =0 \, .
        \label{ClassicalEOM}
    \end{equation}
    By Bianchi identity the term in the parentheses must be constant and we end up with the following,
    \begin{equation}
        R^g_{\mu\nu} - \frac{1}{2} R^g g_{\mu\nu} + \Lambda_{\text{eff}} g_{\mu\nu} =0 \, ,
        \label{ClassicalEOM}
    \end{equation}
    with $\Lambda_\text{eff}$ constant.
    
    \item From the philosophy of our work it follows that the appearance of the crossbreed metric determinants $\sqrt[4]{g\eta}$ is inevitable. Therefore, they cannot be removed by hand, instead, in the quantum regime certain configurations become suppressed and the theory reduces to a scalar-tensor gravity, where the scalar couplings and potentials are obtained from the bi-world principle and are not manifestly imposed. Moreover, since the theory has a specific classical regime, the classical limit and consequently the equilibrium equation is also \textit{obtained} from the same principle. Additionally, because of the specific couplings of $\phi$ with other fields, it plays multiple roles. Within the vacuum, it is a vacuum parameter classically and plays the role of a dilaton in the quantum regime. Above vacuum it couples to matter e.g. in Higgs-like and axion-like ways playing multiple roles. For example, one expects to find in the anomalous non-conservation of the chiral current, terms such as these:
    \begin{equation}
	    \nabla_\mu^g J_5^\mu = \frac{\phi^2}{16 \pi^2 \sqrt{|g|}} \varepsilon^{\mu\nu\rho\sigma} F^\eta_{\mu\nu}F^g_{\rho\sigma} + \frac{\phi^2}{384\pi^2\sqrt{|g|}} \varepsilon^{\mu\nu\rho\sigma} R^{g \ \, \alpha\beta}_{\mu\nu} R^g_{\alpha\beta\rho\sigma} + \dots \, ,
    \end{equation}
    which describes a triangle process with two photons or gravitons on its vertices. So we have a bosonic decay into photons carried by a triangle of fermions (say heavy quarks) and the rate of the decay is determined by how large $\phi$ is.
\end{itemize}

\bibliography{supplement}